\documentclass[10pt,final,journal,a4paper,twoside,twocolumn]{IEEEtran}
\usepackage[T1]{fontenc}
\usepackage{graphicx,latexsym,subfigure,amsmath,epsfig,latexsym,subfigure,amsmath,amssymb}
\usepackage[noadjust]{cite}
\usepackage{algorithm,amsthm}
\usepackage{epstopdf}
\usepackage{cases}
\usepackage{color}
\usepackage{bm}
\usepackage{url}
\usepackage{multirow}
\usepackage{cases}
\usepackage{setspace}
\usepackage{algorithmic}
\usepackage{textcomp}
\usepackage{amsmath}
\usepackage{bm}
\usepackage{algorithm}
\usepackage{algorithmic}
\usepackage{cite}

\begin{document}
\title{Physical Layer Security Enhancement Exploiting Intelligent Reflecting Surface}
\author{\IEEEauthorblockN{Keming Feng, Xiao Li, Yu Han, Shi Jin, and Yijian Chen}\\
\IEEEauthorblockA{National Mobile Communications Research Laboratory, Southeast University\\
Nanjing 210096, China, Email: \{keming\_feng, li\_xiao, hanyu, jinshi\}@seu.edu.cn\\
ZTE Corporation, Shenzhen, China, Email: chen.yijian@zte.com.cn}\\
\thanks{This work has been accepted by IEEE Communications Letters.}
}

\maketitle
\begin{abstract}
In this letter, the use of intelligent reflecting surface (IRS) to enhance the physical layer security of downlink wireless communication is investigated. Assuming a single-antenna legitimate user and a multi-antenna eavesdropper, we propose an effective algorithm to jointly optimize the active and passive beamforming. In the proposed algorithm, the optimal transmit beamforming vector at the BS under fixed IRS phase shifts is derived, and a low-complexity algorithm based on fractional programming (FP) and manifold optimization (MO) is proposed to obtain near optimal IRS phase shifts. Simulation results demonstrate that the proposed algorithm can almost achieve the performance upper bound with a fast convergence rate.
\end{abstract}


\section{Introduction}

Recently, physical layer security (PLS) enhancement has become a non-trivial issue in wireless communication systems due to its capability of replacing traditional cryptography-based techniques \cite{huang2020holographic}. Several endeavors aiming at improving PLS have been proposed, e.g. artificial noise (AN) and cooperative jamming \cite{almohamad2020smart}. However, they both need extra power or hardware costs for practical implementation. Fortunately, with the evolution of metamaterials, there emerges a low-cost and energy efficient device named intelligent reflecting surface (IRS) \cite{almohamad2020smart,huang2019reconfigurable,tang2019TWC,Marco2019EURASIP} offering an effective way to enhance the PLS.

The IRS is composed of reconfigurable, passive, reflecting units with compact size, each of which is able to interact with the incident signal without a dedicated RF processing. By befittingly tuning phase shifts of the reflectors, the reflected signals can be deliberately strengthened or impaired at designated users \cite{feng2020DRL}. Assuming single-antenna eavesdropper, a block coordinate descent (BCD) based algorithm optimizing phase shifts one by one was proposed in \cite{yu2019enabling} for secrecy rate (SR) maximizaiton. The same goal was achieved in \cite{cui2019secure} where the IRS phase shifts were optimized in parallel leveraging Charnes-Cooper transform and semidefinite relaxation (SDR) techniques, which leads to high complexity. 
By introducing AN together with the IRS, joint design of the base station (BS) beamforming vectors, the IRS phase shifts and AN covariance matrixes were investigated in \cite{alexandropoulos2020safeguarding} and \cite{xu2019resource}.

In this letter, with a multi-antenna eavesdropper, we investigate the SR maximization problem with the assist of the IRS. We propose an alternating optimization (AO) algorithm to design the active beamforming at the BS and the phase shifts at the IRS. In the proposed algorithm, the optimal transmit beamforming vector under fixed IRS phase shifts is derived in closed-form. Then, based on fractional programming (FP), the problem of IRS phase shifts optimization is converted into a more tractable form and is conquered iteratively through manifold optimization (MO). Simulation results validate the superiority of the proposed algorithm.


\section{System Model And Problem Formulation}
\begin{figure}[!h]
\centering\includegraphics[width=2.5in]{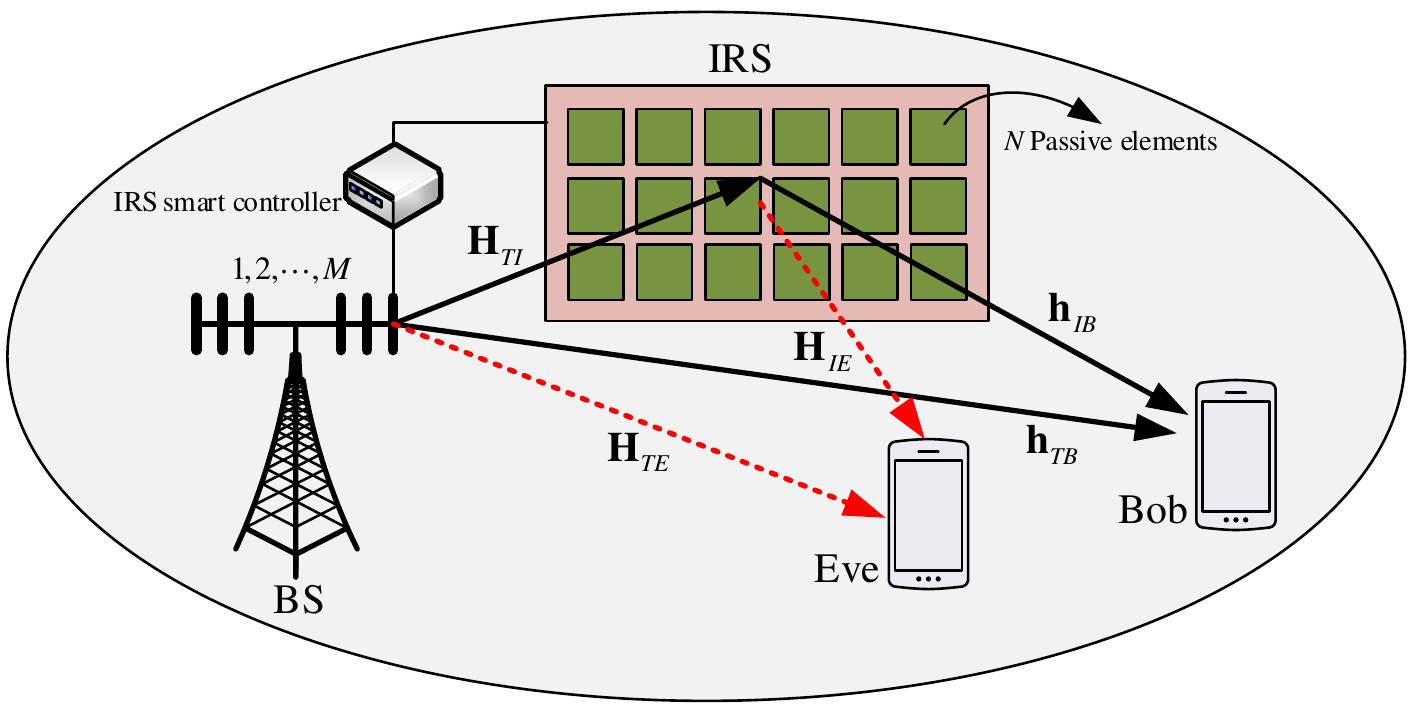}\\
\caption{An IRS-assisted multi-antenna communication system}\label{system}
\end{figure}

Consider an IRS-assisted wiretap communication system consisting of one BS, one legitimate user (Bob), one eavesdropper (Eve), and one IRS, as illustrated in Fig. \ref{system}. The BS is equipped with a $M$-antennas uniform linear array (ULA), Bob is assumed to have a single antenna, while Eve is equipped with $M'$ antennas. An IRS composed of $N$ passive reflectors is deployed to provide a reliable link between the BS and Bob.

The IRS can dynamically adjust the phase shift of each reflecting unit via a smart controller. The baseband equivalent channels between the BS and the IRS, the BS and Bob, the BS and Eve, the IRS and Bob, and the IRS and Eve are represented by $\mathbf{H}_{\textmd{TI}}\in{\mathbb{C}^{N\times{M}}}$, $\mathbf{h}_{\textmd{TB}}\in{\mathbb{C}^{M\times{1}}}$, $\mathbf{H}_{\textmd{TE}}=[\mathbf{h}_{\textmd{TE},1},\cdots,\mathbf{h}_{\textmd{TE},M'}]\in{\mathbb{C}^{M\times{M'}}}$, $\mathbf{h}_{\textmd{IB}}\in{\mathbb{C}^{N\times{1}}}$, $\mathbf{H}_{\textmd{IE}}=[\mathbf{h}_{\textmd{IE},1},\cdots,\mathbf{h}_{\textmd{IE},M'}]\in{\mathbb{C}^{N\times{M'}}}$, respectively. We assume that all channels experience frequency-flat fading and are perfectly known at the BS and IRS.\footnote{Note that the perfect channel state information (CSI) knowledge assumption for both Bob and Eve is reasonable, for example, Eve can also be an active user in the secure transmission system but untrusted by Bob.} Moreover, the power of signals reflected by the IRS two or more times is negligible. Then, the received signals at Bob and Eve can be written as: 
\begin{equation}\label{yb}
  y_\textmd{B} = (\mathbf{h}_{\textmd{IB}}^H\mathbf{\Phi}\mathbf{H}_{\textmd{TI}}+\mathbf{h}_{\textmd{TB}}^H)\mathbf{w}s + n_\textmd{B},
\end{equation}
\begin{equation}\label{ye}
  \mathbf{y}_\textmd{E} = (\mathbf{H}_{\textmd{IE}}^H\mathbf{\Phi}\mathbf{H}_{\textmd{TI}}+\mathbf{H}_{\textmd{TE}}^H)\mathbf{w}s + \mathbf{n}_\textmd{E},
\end{equation}
where $\mathbf{y}_\textmd{E}\in{\mathbb{C}^{M'\times1}}$, $\mathbf{\Phi }=\textmd{diag}\left( {{e}^{j{{\psi }_{1}}}},\cdots, {{e}^{j{{\psi }_{N}}}} \right)$ represents the IRS phase shift matrix, $\psi_i\in{[0,2\pi]}$ denotes the phase shift induced by the $i$-th element on the IRS, $\mathbf{w}\in{\mathbb{C}^{M\times1}}$ is the BS beamforming vector satisfying $\|\mathbf{w}\|^2\leq{P_\textmd{max}}$, $P_\textmd{max}$ is the maximum transmit power, $s$ is the Bob-oriented signal with $\mathbb{E}\{|s|^2\}=1$, $n_\textmd{B}$ and $\mathbf{n}_\textmd{E}$ are additive noises, $n_\textmd{B}$ and each element of $\mathbf{n}_\textmd{E}$ are complex Gaussian variables with zero mean and variances $\sigma_\textmd{B}^2$ and $\sigma_\textmd{E}^2$, respectively. Hence, the achievable rates at Bob and Eve can be specified as
\begin{align}
R_{\textmd{B}} & = \log_2 \left( 1 + \frac{1}{\sigma_\textmd{B}^2}|(\mathbf{h}_{\textmd{IB}}^H\mathbf{\Phi}\mathbf{H}_{\textmd{TI}}+\mathbf{h}_{\textmd{TB}}^H)\mathbf{w}|^2\right),\\
R_{\textmd{E}} & = \log_2 \textmd{det} \left( \mathbf{I}_{M'} + \frac{1}{\sigma_\textmd{E}^2} \mathbf{H}_\textmd{E}^H\mathbf{w}\mathbf{w}^H\mathbf{H}_\textmd{E}\right)\nonumber\\
 & = \log_2 \left( 1 + \frac{1}{\sigma_\textmd{E}^2} \sum \nolimits_{i=1}^{M'}\left|(\mathbf{h}_{\textmd{IE},i}^H \mathbf{\Phi}\mathbf{H}_{\textmd{TI}}+\mathbf{h}_{\textmd{TE},i}^H)\mathbf{w}\right|^2\right),
\end{align}
where $\mathbf{H}_\textmd{E}^H = \mathbf{H}_\textmd{IE}^H\mathbf{\Phi}\mathbf{H}_\textmd{TI}+\mathbf{H}_\textmd{TE}^H$, and the SR from the BS to Bob is \cite{yu2019enabling,cui2019secure}
\begin{equation}
  R_{\textmd{S}}(\mathbf{w},\mathbf{\Phi }) = \left[R_{\textmd{B}}-R_{\textmd{E}}\right]^+,
\end{equation}
where $[x]^+ = \max(0,x)$. Since the maximum SR must be non-negative, the $[\cdot]^+$ operation can be omitted without loss of optimality.

The objective of this paper is to maximize $R_{\textmd{S}}$ by jointly optimizing $\mathbf{w}$ and $\mathbf{\Phi }$, which is formulated as
\begin{align}
 \textmd{(P1):} \quad & \mathop{\max}\limits_{\mathbf{w},\mathbf{\Phi}}\quad
\frac{1+\frac{1}{\sigma_\textmd{B}^2}\left|\left(\mathbf{h}_{\textmd{IB}}^{H}\mathbf{\Phi }\mathbf{H}_{\textmd{TI}}+\mathbf{h}_{\textmd{TB}}^{H}\right)\mathbf{w}\right|^{2}}{1+\frac{1}{\sigma_\textmd{E}^2}\sum\nolimits_{i=1}^{M'}\left|(\mathbf{h}_{\textmd{IE},i}^H\mathbf{\Phi}\mathbf{H}_{\textmd{TI}}+\mathbf{h}_{\textmd{TE},i}^H)\mathbf{w}\right|^2},\label{obj}\\
 \textmd{s.t.} & \quad  {{\left\| \mathbf{w} \right\|}^{2}}\le {{P}_{\max }}, \  \mathbf{\Phi }=\textmd{diag}\left( {{e}^{j{{\psi }_{1}}}},\cdots, {{e}^{j{{\psi }_{N}}}} \right). \label{unit-modulus}
\end{align}
Evidently, (P1) is a NP-hard problem due to the non-convex objective function and unit modulus constraints. Generally, it is impossible to find the optimal solution directly using existed algorithms. Thus, in this paper, we propose a low-complexity AO based algorithm to solve (P1).

\section{Alternating Optimization Based Algorithm}

In this section, we develop an efficient algorithm based on AO to jointly design the transmit beamformer at the BS and the passive phase shift matrix at the IRS. This algorithm works by optimizing $\bm{\Phi}$ and $\mathbf{w}$ alternately with the other one fixed.

First, consider the optimization of $\mathbf{w}$ with fixed $\mathbf{\Phi}$. The optimization problem (P1) degrades to the following problem
\begin{align}
\textmd{(P1a):} \quad &\mathop{\max}\limits_{\mathbf{w}}\quad
\frac{\mathbf{w}^H\mathbf{X}_\textmd{B}\mathbf{w}+1}{\mathbf{w}^H\mathbf{X}_\textmd{E}\mathbf{w}+1}, \label{P1a}\\
& \ \textmd{s.t.} \qquad  {{\left\| \mathbf{w} \right\|}^{2}}\le {{P}_{\max }},
\end{align}
where $\mathbf{X}_\textmd{B} = \frac{1}{\sigma_\textmd{B}^2}(\mathbf{h}_{\textmd{IB}}^H\mathbf{\Phi}\mathbf{H}_{\textmd{TI}}+\mathbf{h}_{\textmd{TB}}^H)^H(\mathbf{h}_{\textmd{IB}}^H\mathbf{\Phi}\mathbf{H}_{\textmd{TI}}+\mathbf{h}_{\textmd{TB}}^H)$ and $\mathbf{X}_\textmd{E} = \frac{1}{\sigma_\textmd{E}^2}\sum\nolimits_{i=1}^{M'}(\mathbf{h}_{\textmd{IE},i}^H\mathbf{\Phi}\mathbf{H}_{\textmd{TI}}+\mathbf{h}_{\textmd{TE},i}^H)^H(\mathbf{h}_{\textmd{IE},i}^H\mathbf{\Phi}\mathbf{H}_{\textmd{TI}}+\mathbf{h}_{\textmd{TE},i}^H)$.

According to \cite{cui2019secure}, the optimal solution to (P1a) is
\begin{equation}\label{w}
  \mathbf{w}^{\star} = \sqrt{P_{\max}}\bm{\lambda}_{\max}(\overline{\mathbf{X}}_{\rm{E}}^{-1}\overline{\mathbf{X}}_{\rm{B}}),
\end{equation}
where $\bm{\lambda}_{\max}(\cdot)$ denotes the normalized eigenvector corresponding to the largest eigenvalue of the input matrix and $\overline{\mathbf{X}}_{i} = \mathbf{X}_{i} + (1/P_{\max}) \mathbf{I}_M, i\in\{\textmd{B,E}\}$.

Then, we try to optimize the phase shift matrix $\mathbf{\Phi}$ by fixing $\mathbf{w}$. The optimization problem becomes
\begin{align}
\textmd{(P1b):} \  & \mathop{\max}\limits_{\mathbf{\Phi}}\
\frac{\left|\left(\mathbf{h}_{\textmd{IB}}^{H}\mathbf{\Phi }\mathbf{H}_{\textmd{TI}}+\mathbf{h}_{\textmd{TB}}^{H}\right)\mathbf{w}\right|^{2}+\sigma_\textmd{B}^2}{\sum\nolimits_{i=1}^{M'}\left|\left(\mathbf{h}_{\textmd{IE},i}^{H}\mathbf{\Phi }\mathbf{H}_{\textmd{TI}}+\mathbf{h}_{\textmd{TE},i}^{H}\right)\mathbf{w}\right|^{2}+\sigma_\textmd{E}^2},\\
& \ \textmd{s.t.}\quad\mathbf{\Phi }=\textmd{diag}\left( {{e}^{j{{\psi }_{1}}}},\cdots, {{e}^{j{{\psi }_{N}}}} \right). \label{unit1}
\end{align}
It is quite hard to derive an optimal solution for (P1b) because of constraint (\ref{unit1}). To make it more tractable, we define $\bm{\theta}=[\theta_1,\cdots, \theta_N]^H=[e^{j\psi_1},\cdots,e^{j\psi_N}]^H$, and reformulate (P1b) as
\begin{align}
\textmd{(P1b'):} \quad&\mathop{\max}\limits_{\bm{\theta}}\quad  f(\bm{\theta}) = \varphi_1(\bm{\theta})+ \varphi_2(\bm{\theta}),\\
& \ \textmd{s.t.} \qquad  |{\theta_i}|=1,\ i=1,2,\cdots,N,
\end{align}
where
\begin{equation}
\varphi_1(\bm{\theta}) = \frac{|\bm{\theta}^H\bm{\alpha}_{\textmd{B}}+
\widetilde{\alpha}_{\textmd{B}}|^2}{\sum\nolimits_{i=1}^{M'}|\bm{\theta}^H\bm{\alpha}_{\textmd{E},i}+\widetilde{\alpha}_{\textmd{E},i}|^2+\sigma_\textmd{E}^2},
\end{equation}
\begin{equation}
\varphi_2(\bm{\theta}) = \frac{\sigma_\textmd{B}^2}{\sum\nolimits_{i=1}^{M'}|\bm{\theta}^H\bm{\alpha}_{\textmd{E},i}+\widetilde{\alpha}_{\textmd{E},i}|^2+\sigma_\textmd{E}^2},
\end{equation}
$\bm{\alpha_{\textmd{B}}}=\textmd{diag}(\mathbf{h}_{\textmd{IB}}^{H})\mathbf{H}_{\textmd{TI}}\mathbf{w}$, $\widetilde{\alpha}_{\textmd{B}} = \mathbf{h}_{\textmd{TB}}^{H}\mathbf{w}$, $\bm{\alpha}_{\textmd{E},i}=\textmd{diag}(\mathbf{h}_{\textmd{IE},i}^{H})\mathbf{H}_{\textmd{TI}}\mathbf{w}$, and $\widetilde{\alpha}_{\textmd{E},i} = \mathbf{h}_{\textmd{TE},i}^{H}\mathbf{w}$.
Note that, (P1b') is a multiple-ratio FP problem \cite{shen2018fractional}, its objective function can be further translated as
\begin{equation}\label{f1}
\begin{split}
&f_1(\bm{\theta}, \mathbf{y}) = 2 \Re \left[ {y_1}^*(\bm{\theta}^H\bm{\alpha}_{\textmd{B}}+\widetilde{\alpha}_{\textmd{B}}) + {y_2}^* \sigma_\textmd{B} \right]\\
&~~~ - \left(|y_1|^2+|y_2|^2\right) \Big(\sum\nolimits_{i=1}^{M'}|\bm{\theta}^H\bm{\alpha}_{\textmd{E},i}+\widetilde{\alpha}_{\textmd{E},i}|^2+\sigma_\textmd{E}^2\Big),\\
\end{split}
\end{equation}
where $\mathbf{y} = [y_1,y_2]^T$ is the introduced auxiliary variable, and $\Re{[x]}$ is the real part of $x$. Then, the equivalent optimization problem becomes
\begin{align}
\textmd{(P1c):} \ &\mathop{\max}\limits_{\bm{\theta},y_1,y_2}\quad
f_1(\bm{\theta},\mathbf{y}),\\
&\ \  \textmd{s.t.} \qquad |{\theta_i}|=1,\ i=1,2,\cdots,N.
\end{align}

Next, we maximize the objective of (P1c) in an iterative fashion over $\bm{\theta}$ and $\mathbf{y}$. Since $f_1(\bm{\theta},\mathbf{y})$ is a quadratic concave function of $\mathbf{y}$ for a given $\bm{\theta}$ \cite{shen2018fractional}, the optimal solutions of the auxiliary variable $\mathbf{y}^{\star} = [y_1^{\star},y_2^{\star}]^T$ under a given $\bm{\theta}$ can be attained by setting $\partial{f_1}/\partial{y_n}=0$, which leads to
\begin{equation}\label{y1}
  y_1^{\star} = \frac{\bm{\theta}^H\bm{\alpha}_{\textmd{B}}+\widetilde{\alpha}_\textmd{B}}{\sum\nolimits_{i=1}^{M'}|\bm{\theta}^H\bm{\alpha}_{\textmd{E},i}+\widetilde{\alpha}_{\textmd{E},i}|^2+\sigma_\textmd{E}^2},
\end{equation}
\begin{equation}\label{y2}
  y_2^{\star} = \frac{\sigma_{\textmd{B}}}{\sum\nolimits_{i=1}^{M'}|\bm{\theta}^H\bm{\alpha}_{\textmd{E},i}+\widetilde{\alpha}_{\textmd{E},i}|^2+\sigma_\textmd{E}^2}.
\end{equation}


Then, under a given $\mathbf{y}$, the objective function with respect to $\bm{\theta}$ can be represented as
\begin{equation}\label{f2}
 \begin{split}
    f_2(\bm{\theta})= f_1(\bm{\theta}, \mathbf{y}) = -\bm{\theta}^H\mathbf{U}\bm{\theta} + 2\Re\left[\bm{\theta}^H\bm{\gamma}\right]+ C,
 \end{split}
\end{equation}
where
\begin{align}
\mathbf{U} & = \left(|y_1|^2+|y_2|^2\right) \sum\nolimits_{i=1}^{M'}\bm{\alpha}_{\textmd{E},i}\bm{\alpha}_{\textmd{E},i}^H,\\
\bm{\gamma} & = {(y_1)}^*\bm{\alpha}_{\textmd{B}} - \left(|y_1|^2+|y_2|^2\right) \sum\nolimits_{i=1}^{M'}\widetilde{\alpha}_{\textmd{E},i}^*\bm{\alpha}_{\textmd{E},i},\\
\label{C}
C & = 2\Re\left[{(y_1)}^*\widetilde{{\alpha}}_{\textmd{B}}+ {(y_2)}^*\sigma_\textmd{B}\right]\nonumber\\
&~~~~ - \left(|y_1|^2+|y_2|^2\right) \Big(\sum\nolimits_{i=1}^{M'}|\widetilde{\alpha}_{\textmd{E},i}|^2+\sigma_\textmd{E}^2\Big).
\end{align}
The phase shift optimization problem can be reformulated as
\begin{align}
\textmd{(P1d):} \quad &\mathop{\max}\limits_{\bm{\theta}} \quad
f_2(\bm{\theta}),\\
&\  \textmd{s.t.} \qquad |{\theta_i}|=1,\ i=1,2,\cdots,N. \label{29}
\end{align}

Note that $\mathbf{U}$ is a positive semidefinite matrix, and $f_2(\bm{\theta})$ is a quadratic concave function of $\bm{\theta}$. Therefore, (P1d) is a quadratically constrained quadratic program (QCQP), and can be solved using the SDR approach. However, the SDR method leads to a computational complexity of $
\mathcal{O}(N^6)$, which is realistically unacceptable. Thus, to reduce the complexity, we strike the non-convexity by invoking principles of MO.\footnote{Note that under the proposed phase shifts optimization framework, (P1d) could also be tackled by the Majorization-Minimization (MM) method \cite{pan2020multicell} whose complexity is similar to the MO approach. Since our goal is to provide an effective solution to the optimization problem and due to the space limitation, we only give the details of the MO method, and provide the performance of the MM method in the simulations.} Note that the feasible search space of the problem in (P1) can be defined as the following complex circle manifold ($\mathcal{CCM}$) 
\begin{equation}
  \mathcal{CCM} = \left\{ \bm{\theta}\in{\mathbb{C}^{N\times{1}}}: |\theta_1|=|\theta_2|\cdots=|\theta_N|=1 \right\},\label{ccm}
\end{equation}
which is also a Riemannian sub-manifold of $\mathbb{C}^N$ \cite{absil2009optimization}, and the unit modulus constraints (\ref{29}) are automatically satisfied when searching the optimal $\bm{\theta}$ over $\mathcal{CCM}$ in each iteration. Therefore, (P1d) can be reformulated as an unconstrained quadratic programming shown below
\begin{align}\label{p1e}
 \textmd{(P1e):}\quad\mathop{\min}\limits_{\bm{\theta}\in{\mathcal{CCM}}}\quad &
f_3(\bm{\theta}) = \bm{\theta}^H\mathbf{U}\bm{\theta} - \bm{\theta}^H\bm{\gamma} -\bm{\gamma}^H\bm{\theta},
\end{align}
where the irrelevant constant term $C$ in (\ref{C}) is ignored.

In the following, we apply the gradient descent framework over $\mathcal{CCM}$ to solve (P1e). In this paper, the conjugate gradient (CG) algorithm is utilized to prevent zigzag phenomenon induced by the steepest gradient descent method. Applying this algorithm, the update rule in the Euclidean space is based on
\begin{equation}\label{update1}
 \bm{x}_{k+1} = \bm{x}_k + \mu_k\bm{\zeta}_k,
\end{equation}
where $ \bm{x}_k \in \mathbb{C}^N$ is the result of the $k$-th iteration in the Euclidean space, $\bm{\zeta}_k \in \mathbb{C}^N$ is the search direction and $\mu_k \in \mathbb{R}$ is the step size. For optimization on $\mathcal{CCM}$,  $\mu_k$ can be determined by the adjusted Armijo-Goldstein condition and the CG descent direction at $\bm{\theta}_{k} \in \mathcal{CCM}$ can be obtained as \cite{chen2017manifold}
\begin{equation}\label{r-update}
  \bm{\zeta}_{k} = -\textmd{grad}_{\bm{\theta}_{k}}^{r}f_3 + \beta_{k-1}\textmd{T}_{{k-1}\rightarrow{{{k}}}}(\bm{\zeta}_{k-1}),
\end{equation}
where $\textmd{grad}_{\bm{\theta}_{k}}^{r}f_3$ is the Riemannian gradient of $f_3(\bm{\theta})$ at $\bm{\theta}_k \in \mathcal{CCM}$ in the tangent space $T_{\bm{\theta}_k}\mathcal{C}$,
\begin{equation}\label{transport}
  \textmd{T}_{{k-1}\rightarrow{{{k}}}}(\bm{\zeta}_{k-1}) = \bm{\zeta}_{k-1} - \Re\left[\bm{\zeta}_{k-1}^*\odot\bm{\theta}_{k}\right]\odot\bm{\theta}_{k},
\end{equation}
$\beta_{k-1}$ is the Polak-Ribi\`{e}re parameter.
According to \cite{chen2017manifold}, the Riemannian gradient can be attained by projecting its Euclidean gradient $\textmd{grad}_{\bm{\theta}_{k}}^{e}f_3$ onto $T_{\bm{\theta}_k}\mathcal{C}$ as
\begin{equation}\label{rgrad}
 \textmd{grad}_{\bm{\theta}_{k}}^{r}f_3 = \textmd{grad}_{\bm{\theta}_{k}}^{e}f_3 - \Re\left[(\textmd{grad}_{\bm{\theta}_{k}}^{e}f_3)^*\odot\bm{\theta}_k\right]\odot\bm{\theta}_k,
\end{equation}
where $\odot$ represents the Hadamard product, and from (\ref{p1e}),  $\textmd{grad}_{\bm{\theta}_{k}}^{e}f_3$ can be calculated as
\begin{equation}\label{egrad}
  \textmd{grad}_{\bm{\theta}_{k}}^{e}f_3 = 2(\mathbf{U}\bm{\theta}_k - \bm{\gamma}). 
\end{equation}
Thus, according to (\ref{update1}), the resulting point of the ($k$+1)-th iteration can be determined by
\begin{equation}\label{theta_next_temp}
  \bm{\theta}_{k+1}' = \bm{\theta}_k + \mu_k\bm{\zeta}_{k}.
\end{equation}
Note that, $\bm{\theta}_{k+1}'$ might not be in ${\mathcal{CCM}}$. The final point $\bm{\theta}_{k+1}$ should be mapped back via \textit{retraction} operation, i.e.,
\begin{equation}\label{theta_next}
  \bm{\theta}_{k+1} = \text{unit}(\bm{\theta}_{k+1}'),
\end{equation}
where $\textmd{unit}(\mathbf{\cdot})$ normalizes all entries of the input vector. The concrete steps of the optimization algorithm solving (P1e) are summarized in Algorithm 1. Since Algorithm 1 is a gradient method with a step size determined by the Armijo-Goldstein rule, it is guaranteed to converge to a stationary point \cite[Theorem 4.3.1]{absil2009optimization}.

\begin{algorithm}[h]
\caption{CG Based Manifold Optimization Algorithm}
\begin{algorithmic}[1]
\STATE Set the iteration index $k = 0$, backtracking scalars $\tau>0$, $\varpi\in(0,1)$, $\alpha\in(0,1)$, pre-defined convergence threshold $\epsilon$, and an initial point $\bm{\theta}_{0}\in\mathcal{CCM}$;
\STATE Calculate $\bm{\zeta}_{k} = -\textmd{grad}_{\bm{\theta}_{k}}^{r}f_3$ based on (\ref{rgrad}) and (\ref{egrad});
\WHILE {True}
 \STATE (Armijo-Goldstein backtracking line search) Find the smallest integer $t\geq0$ satisfies
 $$f_3(\textmd{unit}(\bm{\theta}_k+\tau\alpha^t\bm{\zeta}_{k}))-f_3(\bm{\theta}_k)\leq\varpi\tau\alpha^t\Re[\bm{\zeta}_{k}^H(\textmd{grad}_{\bm{\theta}_{k}}^{r}f_3)]$$
 \STATE Let the line search step size $\mu_k = \tau\alpha^t$;\
 \STATE Determine the next point $\bm{\theta}_{k+1}$ using (\ref{theta_next_temp}) and (\ref{theta_next});\
 \STATE Find the Polak-Ribi\`{e}re parameter $\beta_{k}$ via \cite[Eq. 8.29]{absil2009optimization};\
 \STATE Compute the next search direction $\bm{\zeta}_{k+1}$ with (\ref{r-update}).\
 \IF {$\|\textmd{grad}_{\bm{\theta}_{k+1}}^{r}f_3\|^2 \leq \epsilon$}
  \STATE Set $\bm{\theta}^\star = \bm{\theta}_{k+1}$;
  \STATE \textbf{break};
 \ELSE
  \STATE Set $\bm{\theta}_k = \bm{\theta}_{k+1}$, $\bm{\zeta}_{k} = \bm{\zeta}_{k+1}$, $k = k+1$;
 \ENDIF
\ENDWHILE
\end{algorithmic}
\end{algorithm}

\begin{algorithm}[h]
\caption{IRS phase shifts optimization algorithm}
\begin{algorithmic}[1]
\STATE  Set the iteration number $m = 0$; 
\REPEAT
 \STATE Calculate $y_1^{\star(m)}$, $ y_2^{\star(m)}$ via (\ref{y1}) and (\ref{y2});\
 \STATE Obtain $\bm{\theta}^{\star}$ using Algorithm 1;\
 \STATE Update $m = m+1$, $\bm{\theta}^{(m)} = \bm{\theta}^{\star}$;\
\UNTIL The objective function of (P1b') converges.
\STATE Set the current phase shift matrix $\mathbf{\Phi} = \textmd{diag}((\bm{\theta}^{\star})^*)$.
\end{algorithmic}
\end{algorithm}
The overall optimization algorithm to solve the passive beamforming problem (P1c) is described in Algorithm 2. It can be easily testified that $ f(\bm{\theta}) \geq f_1(\bm{\theta},\mathbf{y})$ with equality holds if $\bf{y} = \bf{y}^\star$. Let $\mathbf{y}^{(q)}$ denote the optimal auxiliary variable determined by (\ref{y1}) and (\ref{y2}) with $\bm{\theta}^{(q)}$, which is the resulting $\bm{\theta}$ of the $q$-th iteration. Obviously, we have
\begin{equation}
\begin{split}
    f(\bm{\theta}^{(q)}) & = f_1(\bm{\theta}^{(q)},\mathbf{y}^{(q)})\overset{(a)}{\leq} f_1(\bm{\theta}^{(q+1)},\mathbf{y}^{(q)}) \\
      & \overset{(b)}{\leq} f_1(\bm{\theta}^{(q+1)},\mathbf{y}^{(q+1)}) = f(\bm{\theta}^{(q+1)}),
\end{split}
\end{equation}
where (a) holds since Algorithm 1 is ensured to converge, (b) holds since the update of $\mathbf{y}^{(q+1)}$ using (\ref{y1}) and (\ref{y2}) maximizes $f_1$ with fixed $\bm{\theta}^{(q+1)}$. Thus, the objective function $f$ is monotonically increasing after each iteration. Since the searching region of $\bm{\theta}$ is bounded, Algorithm 2 is guaranteed to converge.
\begin{algorithm}[h]
\caption{AO Based Algorithm for (P1)}
\begin{algorithmic}[1]
\STATE Set $i = 0$, $\mathbf{w}^{(0)} = \mathbf{h}_{\textmd{TI}}^H/\|\mathbf{h}_{\textmd{TI}}\|$, $\bm{\Phi}^{(0)} = \textmd{diag}(\mathbf{1}_N)$, and ${R_{\textmd{S}}}^{(0)}(\mathbf{w}^{(0)},\mathbf{\Phi}^{(0)})$;
\REPEAT
 \STATE Set $i = i+1$;\
 \STATE With given $\mathbf{\Phi}^{(i-1)}$, update $\mathbf{w}^{(i)}$ with (\ref{w});\
 \STATE Compute $\mathbf{\Phi}^{(i)}$ exploiting Algorithm 2;\
 \STATE Calculate ${R_{\textmd{S}}}^{(i)}(\mathbf{w}^{(i)},\mathbf{\Phi}^{(i)})$;\
\UNTIL {$\left|\frac{{R_{\textmd{S}}}^{(i)}-{R_{\textmd{S}}}^{(i-1)}}{{R_{\textmd{S}}}^{(i)}}\right|\leq \epsilon$};
\STATE \textbf{Return} $\mathbf{\Phi}^{\star} = \mathbf{\Phi}^{(i)} $ and $\mathbf{w}^{\star} = \mathbf{w}^{(i)}$.
\end{algorithmic}
\end{algorithm}

Finally, with Algorithm 2, which optimizes the phase shift matrix under fixed $\bf{w}$, the steps of the AO based algorithm solving (P1) is summarized in Algorithm 3. Similar to Algorithm 2, the SR is monotonically nondecreasing over the iterations and the maximum transmit power is limited. Thus, Algorithm 3 is guaranteed to converge.

Note that the optimization of $\bf{w}$ is matrix inversion involved, thus the computational complexity is $\mathcal{O}(M^3)$. As for $\bm{\Phi}$, the major complexity is built upon the Riemannian search direction update, which is $\mathcal{O}(N^2)$ \cite{chen2017manifold}. Thus, the computational complexity of Algorithm 3 is $\mathcal{O}(I_3(M^3+I_2I_1N^2))$, where $I_1$, $I_2$, and $I_3$ are the iteration indexes of Algorithm 1, 2, and 3, respectively. According to the simulation results, both $I_2$ and $I_3$ are usually less than or equal to 2.

\section{Simulation Results}

We evaluate the performance of the proposed algorithm via numerical results, as compared to the following schemes:
\begin{itemize}
  \item [(1)] Heuristic scheme: it applies the maximum ratio transmission (MRT) at the BS with respect to the direct channel to Bob, and optimizes phase shifts via Algorithm 2.
  \item [(2)] Without-IRS: it straightforwardly optimizes $\mathbf{w}$ via (\ref{w}).
  \item [(3)] CCT-SDR: it optimizes $\bf{\Phi}$ via the Charnes-Cooper transform combined with the SDR approach \cite{cui2019secure}.
  \item [(4)] Proposed-MM: it solves (P1d) via MM method with the other steps in Algorithm 2 remaining unchanged.
  \item [(5)] Upper bound: it is attained by relaxing the rank-one constraint in the semi-definite programming \cite{cui2019secure}.
\end{itemize}
The distance between the BS and the IRS is $d_{\textmd{BI}} = 50$ m, $M = M'= 4$,
$P_{\max}=15\ \textmd{dBm}$, and $\sigma_\textmd{B}^2 = \sigma_\textmd{E}^2 = -75\ \textmd{dBm}$. Both Bob and Eve are located on a line in parallel to that connects the BS and the IRS. The vertical distance between these two lines is $d_v = 2$ m. Assuming all channels follow Rician fading. Specifically, the BS-Bob channel is generated by $\mathbf{h}_{\textmd{TB}} = \sqrt{L_{0}d_{\textmd{TB}}^{-\zeta_{\textmd{TB}}}}\mathbf{g}_{\textmd{TB}}$, where $L_{0} = -30$ \textmd{dB}, $d_{\textmd{TB}}$ is the distance between the BS and Bob, $\zeta_{\textmd{TB}}$ is the path loss exponent, the small-scale fading component
\begin{equation}\label{hTB}
  \mathbf{g}_{\textmd{TB}} =\sqrt{K_{\textmd{TB}}/(K_{\textmd{TB}}+1)}\mathbf{g}_{\textmd{TB}}^{\textmd{LoS}}+ \sqrt{1/(K_{\textmd{TB}}+1)}\mathbf{g}_{\textmd{TB}}^{\textmd{NLoS}},
\end{equation}
$K_{\textmd{TB}}$ is the Rician factor, and $\mathbf{g}_{\textmd{TB}}^{\textmd{LoS}}$ and $\mathbf{g}_{\textmd{TB}}^{\textmd{NLoS}}$ represent the line-of-sight (LoS) and non-LoS (NLoS) components. The channels $\mathbf{H}_{\textmd{TI}}$, $\mathbf{H}_{\textmd{TE}}$, $\mathbf{h}_{\textmd{IB}}$, and $
\mathbf{H}_{\textmd{IE}}
$ adopt the similar channel model. The LoS components are modeled as in \cite{pan2020multicell}. The path loss exponents of the BS-IRS link, the BS-Eve link, the BS-Bob link, the IRS-Eve link, and the IRS-Bob link are set to $\zeta_{\textmd{TI}} = 2.2$, $\zeta_{\textmd{TE}}=3.5$, $\zeta_{\textmd{TB}} = 3.5$, $\zeta_{\textmd{IE}} = 2.5$ and $\zeta_{\textmd{IB}} = 2.5$, respectively. The auxiliary variables introduced in Algorithm 1 are specified as $\tau = 1$, $\varpi = 2^{-13}$ and $\alpha = 0.5$. The convergence threshold is set to $\epsilon=10^{-3}$. The simulation results are averaged over 1000 channel realizations.
\begin{figure}[!h]
\centering\includegraphics[width=2.8in]{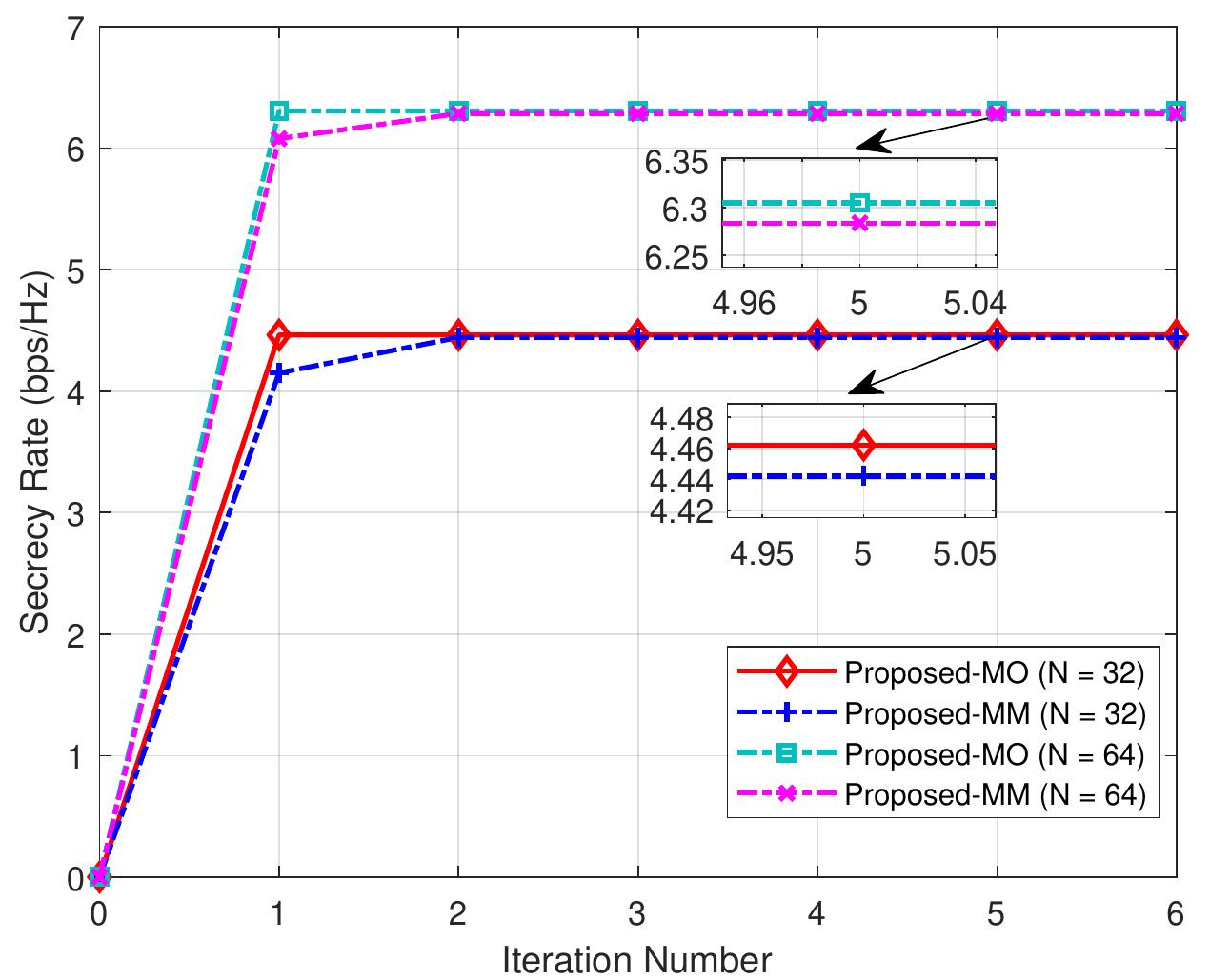}\\
\caption{Convergence Behaviour of the AO Based Algorithm}\label{convergesim}
\end{figure}

We first investigate the convergence behavior of the proposed algorithm (with legend ``Proposed-MO'') in Fig. \ref{convergesim} under two scenarios, i.e., $N = 32$ and $N = 64$, in which the benchmark ``Proposed-MM'' is also compared. We assume that the channels from the BS to Bob and Eve have no LoS components, i.e.,  $K_{\textmd{TB}} = K_{\textmd{TE}} = 0$, and set $K_{\textmd{TI}}=K_{\textmd{IB}}=K_{\textmd{IE}}=10$. The BS-Bob horizontal distance is $d = 48$ m and the BS-Eve horizontal distance is $\tilde{d} = 42$ m. As can be seen, both ``Proposed-MO'' and ``Proposed-MM'' schemes converge in less than 2 iterations, which reveals that the proposed algorithm has a fast convergence rate. Moreover, the performance of the proposed MO-based algorithm is slightly superior to the MM-based algorithm.
\begin{figure}
  \centering
  \includegraphics[width=2.8in]{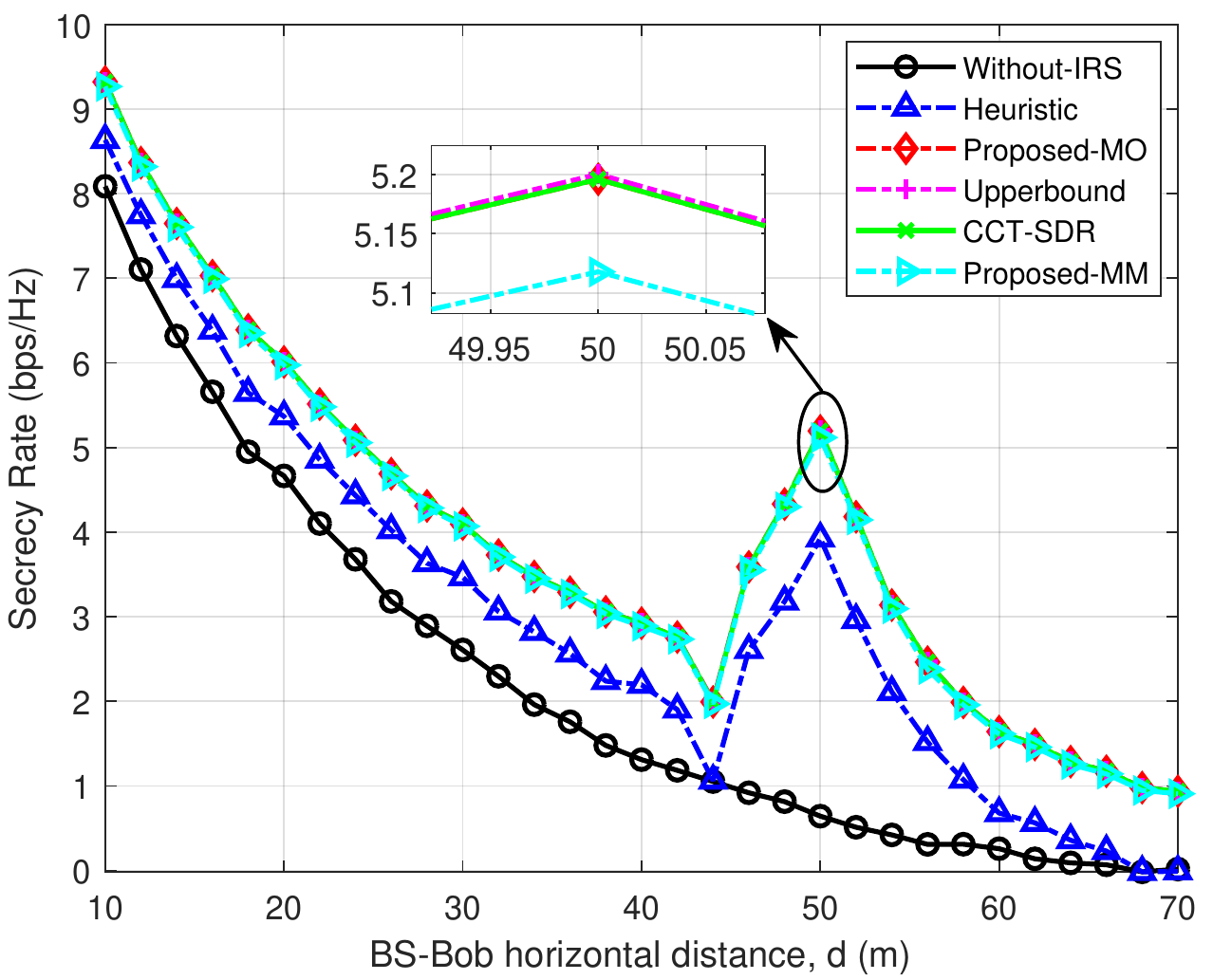}
  \caption{Secrecy Rate vs. The Horizontal Distance between BS and Bob}\label{SR-d-M4}
\end{figure}

Fig. \ref{SR-d-M4} illustrates the achieved SR versus the BS-Bob horizontal distance $d$ when $N=32$ and $\tilde{d} = 44$ m. We set $K_{\textmd{TB}}=K_{\textmd{TE}}=1$, $K_{\textmd{TI}}=10$, and $K_{\textmd{IB}}=K_{\textmd{IE}}=5$. As can be observed, the ``Proposed-MO'' scheme yields performance identical to CCT-SDR (almost achieves the upper bound) and outperforms the other benchmarks. The performance of the heuristic scheme is not satisfactory, since the BS beam is aligned towards Bob only, thus the potential of the IRS is not fully exploited. Note that the SR keeps decreasing when $d$ increases from 10 m to 44 m even with the presence of the IRS, which is expected since and the correlation between the direct links of BS-Bob/Eve becomes stronger \cite{alexandropoulos2017secrecy}. When $d \in [44 \textmd{m},50 \textmd{m}]$, the SR is dramatically improved, which is because: (i) the received siganl is dominated by the reflect link, and the finely designed IRS phase shifts provide another degree of freedom (DoF) rendering higher achievable rate $R_{\textmd{B}}$ as Bob moves close to it \cite{yu2019enabling}; (ii) although Eve is more powerful due to more antennas, the optimal beamforming vector is designed to be in alignment with the legitimate channel while being as orthogonal to the wiretap channel as possible. Hence, little power is received by Eve.
\begin{figure}[!h]
\centering\includegraphics[width=2.8in]{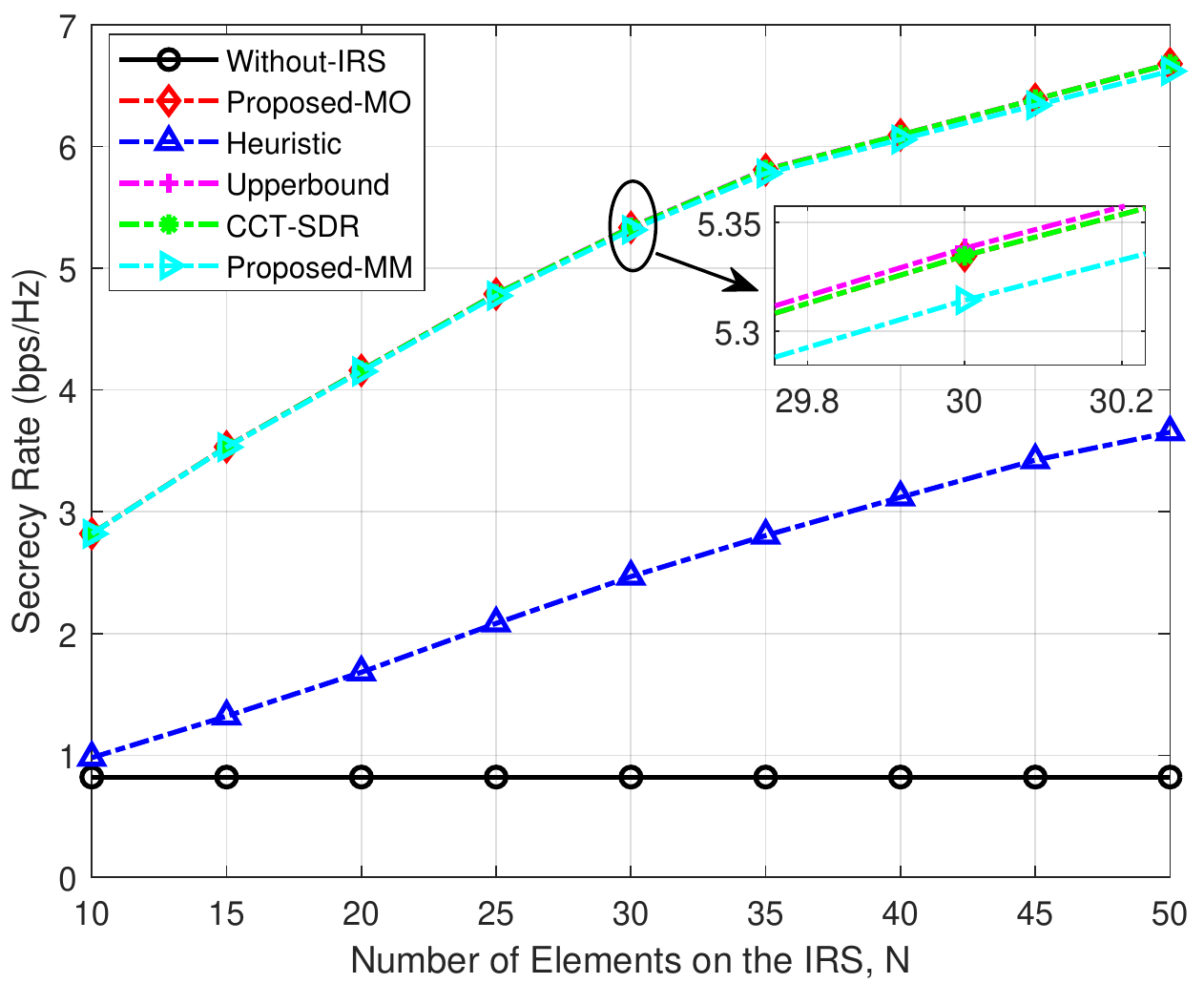}\\
\caption{Secrecy Rate vs. The Number of IRS Reflectors}\label{SR-N}
\end{figure}

Fig. \ref{SR-N} shows the SR versus the number of IRS's reflecting units when $d = 49$ m and $\tilde{d} = 44$ m. The Rician factors are set identical to those in Fig. \ref{convergesim}. It is observed that the SR achieved by the proposed algorithm increases significantly with the increment of $N$ since the signal reflected by the IRS becomes dominant at Bob. It can also be noticed that the performance gap between the proposed algorithm and the heuristic scheme becomes more pronounced due to the joint optimization of $\mathbf{w}$ and $\mathbf{\Phi}$. Note that the performance of the heuristic scheme is similar to that of the ``Without-IRS'' scheme when $N = 10$. This is rational since the received signal is dominated by the direct link other than the IRS-assisted link when $N$ is small. Moreover, the ``Without-IRS'' scheme applies optimal beamforming vector, while the heuristic scheme utilizes sub-optimal beamforming vector, which degrades the performance of the IRS-assisted system.
\begin{table}[]
\centering
\caption{Running time Comparison (in sec.)}\label{table3}
\begin{tabular}{|c|c|c|c|}
\hline
$N$ & \multicolumn{1}{l|}{CCT-SDR} & \multicolumn{1}{l|}{Proposed-MO} &\multicolumn{1}{l|}{Proposed-MM}  \\ \hline
10      & 2.556                              & 0.345                            & 0.331                                      \\ \hline
30      & 4.906                              & 0.621                            & 0.613                                     \\ \hline
50      & 9.603                              & 1.405                            & 1.418                                     \\ \hline
\end{tabular}
\end{table}

Finally, Table \ref{table3} compares the running time of the proposed algorithm with the CCT-SDR algorithm under different $N$ settings. From Table \ref{table3}, it can be seen that the proposed algorithm is much more computationally efficient than the CCT-SDR algorithm. Moreover, the running time of the ``Proposed-MO'' and the ``Proposed-MM'' schemes are almost identical, since their complexity is at the same level \cite{pan2020multicell}.


\section{Conclusion}
This paper investigated the PLS enhancement in an IRS-aided secure transmission system. With the presence of a multi-antenna eavesdropper, the active beamforming at the BS and the passive beamforming at the IRS were jointly optimized for SR maximization. To circumvent the non-convex optimization problem, a low-complexity AO based algorithm exploiting FP and MO techniques was developed. Simulation results validated the superiority of the proposed algorithm in terms of SR and time consumption.

\bibliography{SecrecyRate}
\bibliographystyle{IEEEtran}
\end{document}